# Potential of PEDOT:PSS as a hole selective front contact for silicon heterojunction solar cells

*Sara Jäckle[1,2+], Martin Liebhaber[3+], Clemens Gersmann[3], Mathias Mews[4], Klaus Jäger[5], Silke Christiansen[1,2,6], Klaus Lips[3,6]\**

[+]these authors contributed equally

[1]Staff Department Christiansen, Helmholtz-Zentrum Berlin für Materialien und Energie GmbH, Hahn-Meitner-Platz 1, 14109 Berlin, Germany

[2]Christiansen Research Group, Max-Planck-Institute for the Science of Light, Günther-Scharowsky-Straße 1, 91058 Erlangen, Germany

[3]Institute for Nanospectroscopy, Helmholtz-Zentrum Berlin für Materialien und Energie GmbH, Albert-Einstein-Straße 15, 12489 Berlin, Germany

[4]Institue for Silicon Photovoltaics, Helmholtz-Zentrum Berlin für Materialien und Energie GmbH, Kekuléstraße 5, 12489 Berlin, Germany

[5]Young Investigator Group Nano-SIPPE, Helmholtz-Zentrum Berlin für Materialien und Energie GmbH, Kekuléstraße 5, 12489 Berlin, Germany

[6]Fachbereich Physik, Freie Universität Berlin, Arnimallee 14, 14195 Berlin, Germany

\*lips@helmholtz-berlin.de

## Abstract
We show that the highly conductive polymer poly(3,4-ethylenedioxythiophene)-poly(styrenesulfonate) (PEDOT:PSS) can successfully be applied as a hole selective front contact in silicon heterojunction (SHJ) solar cells. In combination with a superior electron selective heterojunction back contact based on amorphous silicon (a-Si), mono-crystalline n-type silicon (c-Si) solar cells reach power conversion efficiencies up to 14.8% and high open-circuit voltages exceeding 660 mV. Since in the PEDOT:PSS/c-Si/a-Si solar cell the inferior hybrid junction is determining the electrical device performance we are capable of assessing the recombination velocity $v_I$ at the PEDOT:PSS/c-Si interface. An estimated $v_I$ of ~ 400 m/s demonstrates, that while PEDOT:PSS shows an excellent selectivity on n-type c-Si, the passivation quality provided by the formation of a native oxide at the c-Si surface restricts the performance of the hybrid junction. Furthermore, by comparing the measured external quantum efficiency with optical simulations, we quantify the losses due to parasitic absorption of PEDOT:PSS and reflection of the device layer stack. By pointing out ways to better passivate the hybrid interface and to increase the photocurrent we discuss the full potential of PEDOT:PSS as a front contact in SHJ solar cells.

# Introduction

Highly efficient solar cells are often based on heterojunctions, providing charge selective contacts to the absorber material. The world record power conversion efficiency (PCE) of 26.3% for single absorber crystalline silicon (c-Si) solar cells is based on such a heterojunction device structure with hydrogenated intrinsic and doped amorphous silicon (a-Si) contact layers.[1] For c-Si mainly inorganic materials have been investigated as charge selective contacts.[2] The wide range of organic semiconductors, with energy levels often tunable by chemical engineering, promises to find matching charge carrier selective contacts for any possible absorbing semiconductor. Furthermore, in contrast to many inorganic elements and compounds, most organic materials do not require costly vacuum deposition methods and high temperature processes but can cheaply be solution processed involving only low temperature treatments. Especially for the recent interest in combining c-Si with novel solar absorbers like hybrid perovskites in monolithic tandem devices,[3] for which inorganic and organic materials are investigated as contact layers,[4] or with down-converting organic molecule interlayers like tetracene,[5,6] it will be very beneficial to ascertain if organic materials can provide well selective and good passivated junctions for c-Si.

The first promising approach towards hybrid inorganic-organic rectifying junctions on c-Si has been published by Sailor et al. in 1997 using a polymer based on a polyacetylene (PA) backbone with $(CH_3)_3Si$ side groups doped with iodine.[7] In recent years, polythiophenes have been intensively investigated as hole selective contacts for c-Si. Besides poly(3-hexylthiophene) (P3HT),[8] the highly conductive poly(3,4-ethylenedioxythiophene) (PEDOT) in a complex with poly(styrene sulfonate) (PSS) shows promising results on n-type c-Si. First PEDOT:PSS/c-Si devices have been presented in 2010,[9] followed by numerous reports on modifications of this junction: for example by either nanostructuring the c-Si at the interface[10–12] or implementing an inorganic[13] or organic interlayer.[14] In all of these studies the performance of the solar cells was actually not limited by the hybrid interface itself but rather by the intrinsic properties of the c-Si wafer or the implemented back contact. For instance, if the minority carrier diffusion length is shorter than the wafer thickness, the open-circuit voltage ($V_{OC}$) and hence the efficiency is determined by the bulk lifetime of the c-Si absorber.[15,16] When this is not the case, the mostly used ohmic metallic back contact is limiting the solar cells by its high recombination velocity. Schmidt et al. combined the PEDOT:PSS front junction with a conventional diffused back surface field (homojunction) on a moderately doped high quality c-Si wafer.[17] These PEDOT:PSS/c-Si/c-Si($n^+$) devices reached efficiencies up to 12.7% with a $V_{OC}$ of 603 mV. Only when inverting the device structure, using PEDOT:PSS as a back junction and standard optimization of the front side as in conventional homojunction c-Si solar cells, a $V_{OC}$ of 657 mV and an efficiency up to 20.6% could be reached.[18] Combining the hybrid PEDOT:PSS/c-Si front junction with a hetero back junction, was first attempted in 2014 by Zhang et al.[19] Inserting a cesium carbonate ($Cs_2CO_3$) layer between c-Si and the rear metal electrode led to an increase in $V_{OC}$ of almost 40 mV to 621 mV. Namagatsu et al. used titanium dioxide ($TiO_2$) as an electron selective back contact resulting in efficiencies up to 11.2% with a $V_{OC}$ of 614 mV.[20] Nevertheless in this case the device is still limited by the $TiO_2$/c-Si back junction.

In the present study, we investigate the limitations of PEDOT:PSS as a hole selective front junction in hybrid SHJ solar cells by combing it with a superior, highly electron selective and well passivated, back contact on a Si wafer with a long bulk lifetime. For this we deposit a layer stack of intrinsic and $n^+$-doped hydrogenated amorphous silicon (a-Si:H), adopted from the high efficiency SHJ technology, on the backside of the wafer. The performance of the PEDOT:PSS/c-Si/a-Si solar cells is assessed by illuminated current density-voltage (*J-V*) and small signal capacitance-voltage (*C-V*) measurements. The selectivity and the passivation of the hybrid PEDOT:PSS/c-Si junction will be discussed by estimating the interface recombination velocity. Finally, we address the major optical losses and discuss the potential of the hybrid device concept by comparing the external quantum efficiency (EQE) with optical simulations of the layer stack.

## Methods
**Preparation of PEDOT:PSS/c-Si/metal solar cells**
PEDOT:PSS/c-Si/metal solar cells were prepared on a 4" Si wafer (mono-crystalline, n-type, Si(100), thickness 525 μm, resistivity 1 – 5 Ωcm) cut into 1.5 × 1.5 cm² samples and cleaned by ultrasonication in acetone and isopropanol. To define and isolate an area of 1.17 cm², a photoresist (nLof, Microchemicals) was spin-coated onto the samples and developed by UV lithography. The native oxide on the c-Si surface was removed by dipping in hydrofluoric acid (5% HF) for 30 s, followed by a subsequent dip in deionized (DI) water. PEDOT:PSS (PH1000, Heraeus Clevios) was filtered with a polyvinylidene fluoride membrane (0.45 μm porosity) to remove agglomerates. To the solution 5 vol% dimethyl sulfoxide (DMSO) and a wetting agent (0.1 vol% FS31, Capstone) were added. The polymer was spin-coated onto the c-Si substrates at 4000 rpm for 20 s and subsequently annealed at 130 °C for 15 min in ambient air. To reduce edge effects, the area defined by lithography was masked and the surrounding polymer was etched away by a strong oxygen plasma. For complete photovoltaic devices an In/Ga eutectic was scratched into the c-Si as a metal back contact and a 300 nm Au grid (finger width 80 μm) was evaporated by an electron beam through a shadow mask onto the polymer as a front electrode, leaving an active area of 0.82 cm². For electrical measurements the solar cells were placed on a copper tape.

**Preparation of PEDOT:PSS/c-Si/a-Si heterojunction solar cells**
The a-Si:H/c-Si heterojunction back contact was fabricated on a (both side) polished 4" Si wafer (float zone grown, mono-crystalline, n-type, Si(100), thickness 280 μm, resistivity 1 – 5 Ωcm) prior to polymer processing. After standard RCA cleaning and removal of the native oxide layer by a HF dip (1%, 2 min), an amorphous Si layer stack consisting of a 4 nm-thick intrinsic a-Si:H followed by a 8 nm-thick highly n-doped a-Si:H was grown by plasma enhanced chemical vapor deposition. Subsequently an 80 nm thick indium-tin-oxide (ITO) layer was deposited by RF magnetron sputtering and the stack was annealed at 200 °C for 5 min. The back contact of the solar cell was completed by metallization with a 10 nm thermally evaporated Ti adhesion layer and 500 nm Ag. Details about the Ag/Ti/ITO/a-Si:H(n)/a-Si:H(i)/c-Si electron back contact can be found elsewhere.[21] After back contact processing the wafer was cut into 2.5 × 2.5 cm² large substrates. Prior to polymer deposition, the native oxide on the front side of the c-Si substrates was removed by a HF drop (1%, 2 min), sparing the back side, followed by a dip in DI water. PEDOT:PSS (F HC, Heraeus Clevios) was then spin-coated at 2000 rpm for 20 s onto the c-Si substrates and subsequently annealed at 130°C for 15 min in ambient air. The polymer solutions PH1000 and F HC can be used interchangeably when controlling the polymer thickness (see Supplementary Information S1). The solar cell fabrication was completed by thermal evaporation of a 300 nm-thick Ag front grid (finger width 200 μm). Finally, the substrate was cut into 1 × 1 cm² solar cells with an active area of 0.85 cm². A schematic of the device structure is displayed in Figure 1. For electrical measurements the solar cells were glued with a conductive epoxy to a copper sheet.

**Electrical device characterization**
Electrical device characterization was carried out using a four-terminal contact configuration. To characterize the photovoltaic response of the devices, samples were irradiated through the transparent PEDOT:PSS layer with an AM1.5G reference spectrum (Class AAA Solar Simulator) and current density-voltage (*J-V*) curves were measured. All standard solar cell parameters were derived from these illuminated *J-V* curves. Capacitance-voltage (*C-V*) measurements were carried out using a Keithley SCS 4200 measure unit operated at 10 kHz with an AC amplitude of 10 mV and a voltage sweep between −2 V and +2 V. The built-in voltage $\psi_{bi}$ of the hybrid c-Si(n)/PEDOT:PSS junction as well as the doping density $N_D$ of the Si wafer were obtained from the *V*-axis intercept and the slope of the linearly fitted data, respectively.[16] The external quantum efficiency (EQE) was measured with light from a 300 W Xenon source coupled through a CS260 monochromator (Newport) and focused on a spot small enough to fit in-between grid lines. A c-Si reference cell was used for calibration.

**Optical simulations**
Optical simulations were performed with the MATLAB software package 'GenPro4', which was developed at Delft University of Technology and is based on a transfer-matrix algorithm.[22] GenPro4 allows to distinguish between coherent and incoherent layers and calculates the absorption spectra $A_i(\lambda)$ of each layer as well as the reflection of the layer stack. The current density $J_i$ corresponding directly to the absorption in the layer *i* assuming an IQE = 1, is then calculated using

$$J_i = -e \int_{300 \text{ nm}}^{1180 \text{ nm}} A_i(\lambda)\, \Phi_{\text{AM1.5}}(\lambda)\, \mathrm{d}\lambda$$

where $\Phi_{\text{AM1.5}}$ denotes the photon flux according to the AM1.5G spectrum.[23] For the simulations of the PEDOT:PSS/c-Si/a-Si layer stack we assumed planar interfaces. The 280 μm thick c-Si wafer can be treated as an incoherent layer.[22] As input for the simulations, experimentally determined complex refractive indices (*n-k*) of the materials are used. The complex refractive index of the PEDOT:PSS formulation (F HC) used in this work was derived by spectroscopic ellipsometry (UVISEL, Horiba Jobin Yvon). The measurement was performed for a photon energy range from 0.6 eV - 4.8 eV with steps of 0.02 eV under an angle of incidence of 70°. The data was analyzed assuming uniaxial anisotropy and including the Drude model and a Lorentz oscillator for each component, following Pettersson et al.[24] (see Supplementary Information S2). For the simulations only the ordinary component has to be considered because of the planar interfaces and the normal incidence of light.

# Results and Discussion
PEDOT:PSS is applied as a hole selective front contact on Si wafers of different quality featuring different back contacts. Bulk-limited PEDOT:PSS/c-Si solar cells are fabricated by using a 525 μm thick Si wafer with a low minority carrier diffusion length of ~ 260 μm ($L \ll d$)[16] and a metal In/Ga eutectic back contact. Contact-limited PEDOT:PSS/c-Si solar cells use a ~ 280 μm thick high quality Si wafer with a long minority carrier diffusion length ($L \gg d$). As a back contact either an electron selective a-Si(i)/a-Si(n$^+$)/ITO/Ti/Ag layer stack is deposited or, for comparison, a metal In/Ga eutectic back contact is applied. The photovoltaic response of the different solar cells is displayed in Figure 2a and the characteristic device parameters are collected in Table 1. The bulk-limited solar cell with the lower quality Si wafer ($L \ll d$) has a $V_{OC}$ of 544 mV and a PCE of 12.2%. This compares well to earlier results, where we showed in detail that for the here used Si wafer device performance depends on c-Si bulk lifetime and not on the hybrid front junction, proving that PEDOT:PSS/c-Si cannot be considered as a metal-semiconductor junction.[16] Using instead a high quality Si wafer with a long bulk lifetime ($L \gg d$) ensures that the solar cells will be limited by the recombination at the interfaces and the selectivity of both contacts. The photovoltaic response of the contact-limited hybrid solar cells featuring an ohmic metal back junction is not much different from that of the bulk-limited device. It exhibits only a slightly larger short-circuit current density ($J_{SC}$) with a similar $V_{OC}$ of 550 mV and PCE of 12.9%. Because of the long lifetime all holes (minority carriers in n-type Si), which are generated within the c-Si absorber, can diffuse to the back side. As a metal has in principle an infinitely large recombination velocity, due to a continuous density of states,[25] all holes reaching the ohmic backside will recombine there. As the thickness of the high quality wafers (~ 280 μm) used for this metal contact-limited device is only slightly higher than the minority carrier diffusion length in the low quality wafers used for the bulk-limited device (~ 260 μm[16]) this leads to almost identical performance of the solar cells. The contact-limited device featuring an a-Si(i)/a-Si(n$^+$)/ITO/Ti/Ag layer stack as an electron selective back junction, as depicted in Figure. 1, shows a clearly improved $V_{OC}$ of 663 mV leading to a PCE of 14.8%. To the best of the authors knowledge this is the highest $V_{OC}$ reported to date for a c-Si based solar cell featuring PEDOT:PSS as a front contact. All-inorganic SHJ solar cells using the same a-Si layer stack as a back contact on similar Si wafers and featuring Ti/Ag-front grid/ITO/ a-Si(p$^+$)/a-Si(i) as a front contact lead to $V_{OC}$s well above 700 mV[21] and PCEs above 16%[26], clearly pointing out that in our case the solar cell is limited by the hybrid PEDOT:PSS/c-Si front junction. This is also supported by a similarly high $V_{OC}$ of 657 mV that has been observed using PEDOT:PSS as a hole selective back contact for highly efficient Si solar cells with a front homojunction.[18] Even though PEDOT:PSS does not perform as well as an a-Si front junction, it shows the best performance off all organic materials tested so far. Using a similar a-Si layer stack as an electron selective back contact with P3HT as a front contact leads only to a $V_{OC}$ of 617 mV.[27]

In order to understand the limiting properties of the hybrid PEDOT:PSS/c-Si front junction in our device, we take into account the electronic structure derived from previous work as well as values obtained by capacitance measurements. $N_D$ and $\psi_{bi}$ of the bulk-limited as well as the contact-limited PEDOT:PSS/c-Si solar cells are derived from Mott-Schottky plots displayed in Figure 2b and the values are collected in Table 1. As expected, $N_D$ of the Si wafer in both devices is similar. Together with the valence and conduction band position in c-Si, as taken from literature,[28] the Fermi level position can be determined to be at 4.3 eV. Independent of the device concept used, $\psi_{bi}$ is almost identical being slightly above 690 mV. The extracted $\psi_{bi}$ of the PEDOT:PSS/c-Si solar cells is only about 30 mV larger than the measured $V_{OC}$ of our PEDOT:PSS/c-Si/a-Si solar cell. With the Fermi level position in c-Si and the extracted $\psi_{bi}$, assuming no Fermi level pinning,[16] the work function of PEDOT:PSS is ~ 5.0 eV. The valence band / HOMO position of the polymer can be estimated to be at 4.9 eV from its relative position to the Fermi level, as previously determined.[16] The position of the conduction band / LUMO was recently measured by Inverse Photoemission Spectroscopy to be at 3.6 eV.[20] Including also the band structure and Fermi level position for thin n$^+$-doped a-Si:H layers,[29] leads to the schematic band diagram for the PEDOT:PSS/c-Si(n)/a-Si solar cell, depicted in Figure 3 in the dark.

A solar cell junction is basically characterized by two parameters, its selectivity and its passivation. The selectivity is given by the potential barrier for one type of charge carriers. In a heterojunction this barrier consists of the build-in potential $\psi_{bi}$, the difference between the Fermi level position in the two semiconductors, and additionally the band offset $\Delta E_C$ (cf. Fig. 3). The passivation is characterized by the surface recombination velocity $v_S$ at the metal front electrode and, in heterojunctions more importantly, the interface recombination velocity $v_I$. This leads to the following contribution of the junction to the dark recombination current density $J_0$ of the heterojunction solar cell:

$$J_0 \approx qN_D v_I e^{-\frac{q\Psi_{bi}}{kT}} + qN_D v_S e^{-\frac{q(\Psi_{bi}+\Delta E_C)}{kT}} \quad \begin{array}{l} \textit{selectivity} \\ \textit{passivation} \end{array} \quad (1)$$

All carriers $qN_D$ that pass over the build-in potential with the probability $e^{-\frac{q\Psi_{bi}}{kT}}$ recombine with the velocity $v_I$ at the interface while all carriers that pass over the complete barrier recombine with $v_S$ at the surface. Extracted from capacitance measurements, PEDOT:PSS introduces a $\psi_{bi}$ of 690 meV in c-Si. Additionally, it features a conduction band offset $\Delta E_C$ of 0.45 eV to c-Si. Together resulting in the good hole selectivity comparably to p-doped a-Si:H with a valence band offset of 0.44 eV.[29] Accounting for the lower $V_{OC}$ of the hybrid PEDOT:PSS/c-Si/a-Si solar cells, leads to the assumption that the c-Si surface in the PEDOT:PSS/c-Si junction cannot be as well passivated as in the conventional c-Si/a-Si junction by the thin a-Si:H(i) interlayer. The dark recombination current density $J_0$ of our solar cell can be approximated with the measured $V_{OC}$ and $J_{SC}$ by the ideal diode equation at open-circuit:[30]

$$J \approx J_0 \left[\exp\left(\frac{qV}{kT}\right) - 1\right] - J_{sc} \xrightarrow{\text{open circuit } J=0} J_o \approx \frac{J_{sc}}{exp\left(\frac{qV_{oc}}{kT}\right) - 1} = 2 \times 10^{-13} \frac{A}{cm^2}$$

By assuming that $J_0$ in our contact-limited PEDOT:PSS/c-Si/a-Si solar cells is only determined by the hybrid front junction, we can estimate the PEDOT:PSS/c-Si interface recombination velocity $v_I$ from Equation 1. For the surface recombination velocity $v_S$ thermionic recombination at a metal-semiconductor interface is considered, so all electrons that can surpass the barrier recombine with a velocity of $v_S \approx A^{**}T/qN_C$ ($A^{**}$ denoting the reduced effective Richardson constant).[28] This leads to an interface recombination velocity extracted from Equation 1 of:

$$v_I \approx 365 \ \frac{cm}{s}$$

The estimated interface recombination velocity of about 400 cm/s fits well to the observation that the Si wafer actually exhibits a native oxide at the interface to PEDOT:PSS, as we have shown previously.[31] The recombination velocity at the surface of a n-type Si wafer with a native oxide is usually in the order of 1000 cm/s, strongly depending on the time after HF treatment.[32] The here estimated dark emitter recombination current density of $\sim 2 \times 10^{-13}$ A/cm$^2$, is slightly higher than the one measured by Schmidt et al. with transient photoconductance decay for a PEDOT:PSS/SiO$_x$/c-Si interface.[17] This might be explained by the fact that we neglect any recombination channels besides the interface itself. Another possibility is that a different state of native oxide formation at the interface was measured, as this is not stable.[31]

To assess the limitations and the potential of PEDOT:PSS as a front contact from an optical point of view, we performed optical simulations of the PEDOT:PSS/c-Si/a-Si layer stack. In Figure 4a the corresponding maximal expected short-circuit current density (grey line), as well as current density losses due to parasitic absorption of the polymer and reflection of the layer stack as a function of PEDOT:PSS thickness are presented. While including the back junction and metallization into the optical simulations is important to account for back reflection in the long wavelength region (> 900 nm), there is almost no parasitic absorption in these layers. Comparing the measured external quantum efficiency (EQE) of the PEDOT:PSS/c-Si/a-Si solar cells to the absorption in the Si wafer at various polymer layer thicknesses by optical simulations, we predict a polymer thickness of 95 nm in the here presented solar cell. This is in good agreement with the measured thickness of ~ 93 nm at the center of the device from atomic force microscopy. Figure 4b shows the measured EQE of the contact-limited PEDOT:PSS/c-Si/a-Si solar cell together with the simulated absorption in the Si wafer and the polymer as well as the reflection of this device. As indicated by the vertical dashed line in Figure 4a, at this PEDOT:PSS thickness about 6 mA/cm$^2$ are lost due to parasitic absorption of the polymer and about 10 mA/cm$^2$ to reflection, thus limiting the maximal possible $J_{SC}$ to ~ 30 mA/cm$^2$. The high parasitic absorption losses could easily be reduced by using a thinner polymer while at the same time adding a non-absorbing layer with a similar refractive index, like lithium fluoride LiF or molybdenum oxide $MoO_x$[33] to maintain low reflection. We have shown previously that it is possible to reduce the polymer thickness by ~ 50% while maintaining the same sheet resistance through a solvent post treatment.[31] This could reduce the current loss due to parasitic absorption down to 2 mA/cm$^2$, increasing the PCE to above 16%. The reflection losses mainly origin from internal reflection at the planar interface between PEDOT:PSS and c-Si due to their large difference in refractive indices. This can be decreased by structuring the interface. At the same time structuring would also reduce the parasitic absorption of the polymer further, as the part of the light hitting the tilted surface would oscillate in direction of the polymers extraordinary component that shows a drastically reduced extinction coefficient (see Supplementary Information S2). This is most likely due to the free carriers that can only move in direction of the polymer chains which are aligned parallel to the surface.[24] Standard KOH etching of c-Si(100) wafers results in pyramidal structured surfaces with smooth {111} planes tilted 54.7° with respect to the {100} plane.[34] This would lead to a reduction of the parasitic absorption by about 40%. However polymer deposition on structured c-Si surfaces has been shown difficult so far and was therefore always accompanied by a loss in $V_{OC}$.[10,12,17] Also for incomplete coverage on structured c-Si a stable and well passivating layer at the interface to PEDOT:PSS might be beneficial. This could possibly either be accomplished by an organic functionalization or an inorganic interlayer, i.e. a thin, well defined and stable chemical tunnel oxide grown on the c-Si surface prior to polymer deposition. Such an oxide has shown excellent passivation properties in all-inorganic c-Si solar cells.[35]

## Conclusion and Outlook

By combing a PEDOT:PSS front junction on c-Si with a superior a-Si back contact, we present solar cells with an efficiency of 14.8% and an open-circuit voltage $V_{OC}$ exceeding 660 mV. This high $V_{OC}$ is the highest value achieved for c-Si based solar cells using PEDOT:PSS as a front junction, and matches well the highest $V_{OC}$ reported in literature using PEDOT:PSS as a back junction in combination with an optimized diffused front homojunction.[18] A comparison of the performance of the presented PEDOT:PSS/c-Si/a-Si solar cell with similarly prepared complete inorganic c-Si/a-Si heterojunction solar cells reveals that the here presented device is limited by the hybrid PEDOT:PSS/c-Si interface. This justifies the assumption that the dark recombination current of the presented solar cell arises mostly from the hybrid junction. By extracting the built-in voltage through capacitance measurements, we show that the junction is well selective and furthermore we are able to approximate the PEDOT:PSS/c-Si interface recombination velocity. The recombination velocity of ~ 400 m/s corresponds well to the observation that the c-Si surface is passivated by a native oxide at the interface to the polymer under standard fabrication. With this we are able to point out that the hybrid PEDOT:PSS/c-Si junction is limited by the interface recombination velocity and the $V_{OC}$ of hybrid SHJ solar cells could be further improved by providing a better passivating and stable interlayer.

Comparing the measured external quantum efficiency of the PEDOT:PSS/c-Si/a-Si solar cells with optical simulations reveals that about 6 mA/cm$^2$ of the short-circuit current density is lost because of the parasitic absorption of PEDOT:PSS and about 10 mA/cm$^2$ due to reflection of the planar layer stack. By a reduction of the polymer thickness efficiencies above 16% should be possible. A further increase in current would have to be accompanied by surface structuring of the c-Si wafer, leading to a lower effective extinction coefficient and multiple incidents of light. As full-coverage of the surface by polymer deposition has been shown difficult, also in this case a better passivation concepts of the interface might be beneficial.

The example of PEDOT:PSS shows that organic materials can very well provide charge selective junctions for inorganic semiconductors like silicon, pioneering many possible new hybrid approaches for solar cells.

## Acknowledgements
S.J. and S.C. recognize the financial support by the SFB 951 "Hybrid Inorganic/Organic Systems (HIOS) for Opto-Electronics". M.L. and K.L. like to acknowledge financial support provided by the German Federal Ministry for Research and Education (BMBF) through the project "Silicon In-situ Spectroscopy at the Synchrotron" (SISSY), Grant No. BMBF03SF0403.

## Author Contributions
S.J. and M.L. equally contributed to this work. M.M. fabricated the electron selective back contact, i.e. PECVD grown amorphous Si layer stacks, ITO sputter deposition and metallization. The fabrication of the hybrid front contacts as well as device characterization, i.e. *J-V* characteristics, *C-V* response and EQE measurements were done by S.J., C.G. and M.L. Optical constants of PEDOT:PSS were determined by spectral ellipsometry by S.J. Using the software GenPro4, optical simulations were carried out by M.L. and K.J. The manuscript was prepared under the responsibility of S.J., edited and discussed by all authors.

## Additional information
Supplementary information is available online.

## Competing financial interests
The authors declare no competing financial interests.

# Figures

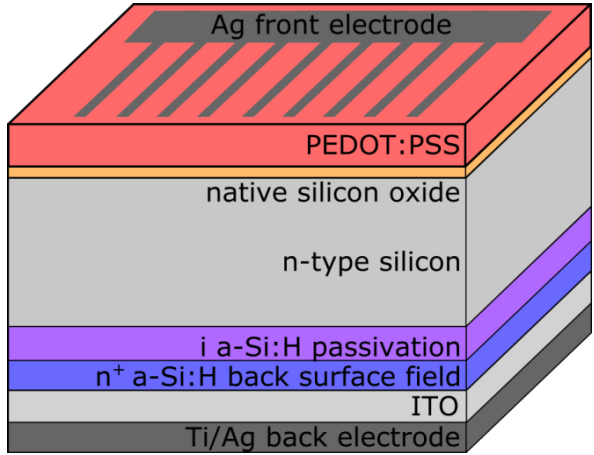

**Figure 1. Schematic of the device structure of the PEDOT:PSS/c-Si/a-Si solar cells.** On the backside of a mono-crystalline n-type Si wafer an intrinsic and a highly n-doped amorphous silicon layer as well as an ITO layer and Ti/Ag electrode are deposited. On the front side PEDOT:PSS is spin coated and as a front electrode an Ag-grid is deposited.

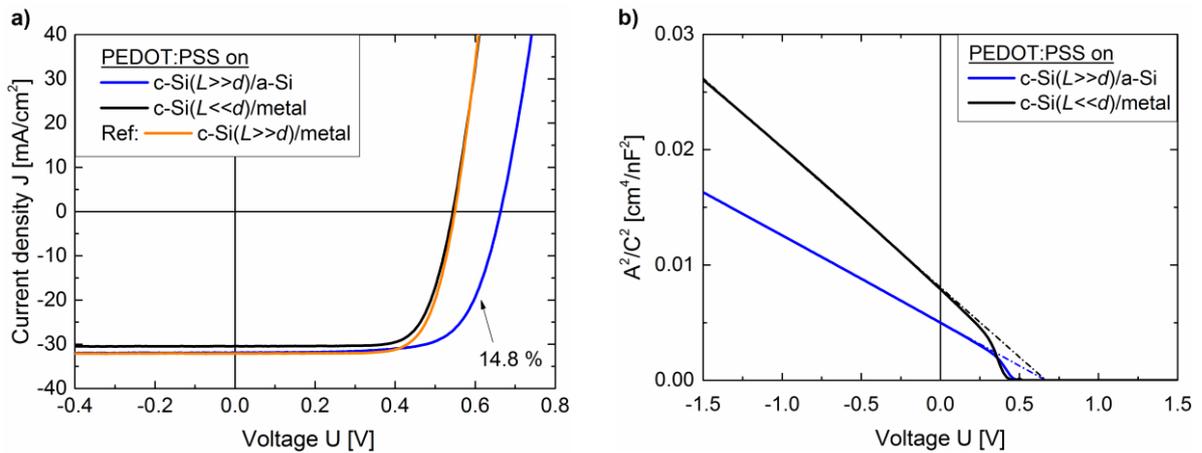

**Figure 2. a) Photovoltaic (*J-V*) and b) Capacitance (*C-V*) response of c-Si based solar cells with a PEDOT:PSS front contact.** Comparing PEDOT:PSS/c-Si/a-Si(i)/a-Si($n^+$)/ITO/Ti/Ag solar cells based on Si wafers with a low minority carrier diffusion length $L$ compared to its thickness $d$ ($L \gg d$) to PEDOT:PSS/Si/metal solar cells based on lower quality Si wafer ($L \ll d$). As a reference it is shown that the a-Si back contact does not lead to an increase in performance of the hybrid solar cell when applied to the lower quality wafer. All extracted solar cell parameters are collected in Table 1.

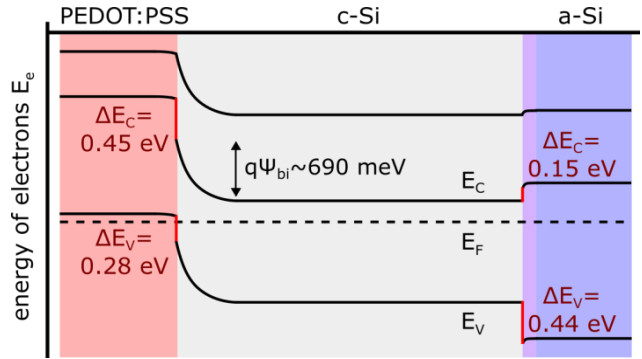

**Figure 3. Schematic band diagram of a heterojunction solar cell with a hybrid PEDOT:PSS/c-Si front junction and a c-Si/a-Si(i)/a-Si(n$^+$) back junction in the dark.** The built-in voltage $\psi_{bi}$ is obtained by *C-V* measurements. Conduction and valence band positions are taken from literature. The valence and conduction band offsets are marked red.

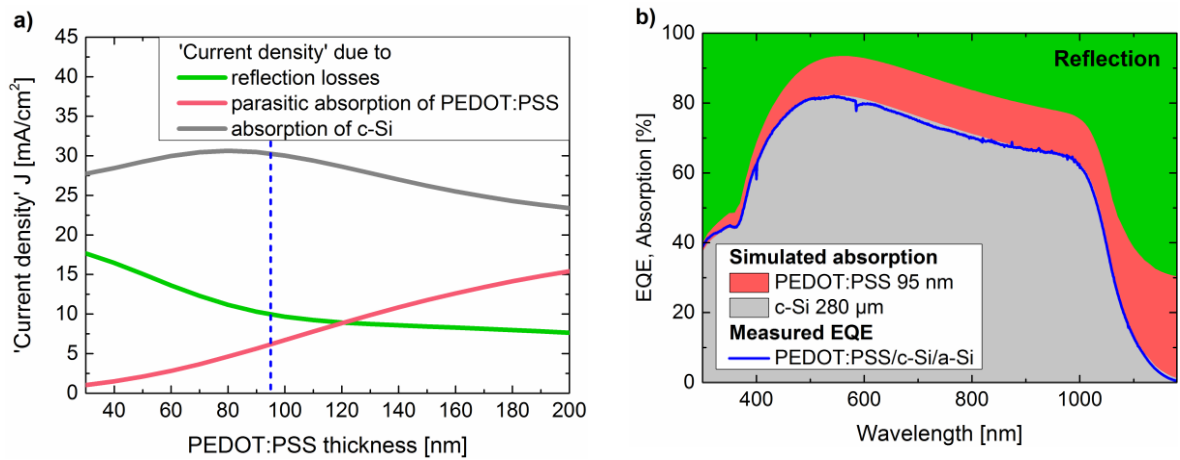

**Figure 4. Simulation of the optical response of the PEDOT:PSS/c-Si/a-Si layer stack. a)** Simulated maximal current densities expected due to the absorption in the c-Si wafer as well as due to losses by reflection and by parasitic absorption of PEDOT:PSS for varying polymer thicknesses. **b)** Reflection and absorption in a solar cell with 95 nm of PEDOT:PSS compared to the measured EQE spectrum of the here presented PEDOT:PSS/c-Si/a-Si solar cell.

# Table

**Table 1. Summary of the measured PEDOT:PSS/c-Si solar cell device parameters.** The doping density $N_D$ of the c-Si wafers and the build-in voltage $\psi_{bi}$ in the device are derived from the *C-V* measurements in Fig. 2b. The open-circuit voltage $V_{OC}$, short-circuit current density $J_{SC}$, fill factor FF, and the power conversion efficiency PCE of the solar cells are extracted from the illuminated *J-V* curves in Fig. 2a.

| PEDOT:PSS on: | Capacitance | | Illuminated J-V curve | | | |
|---|---|---|---|---|---|---|
| | $N_D$ [cm$^{-3}$] | $\psi_{bi}$ [V] | $V_{oc}$ [V] | $J_{sc}$ [mA/cm²] | FF | PCE [%] |
| **c-Si(*L << d*)/metal** | 1.0 x 10$^{15}$ | 0.692 | 0.544 | 30.5 | 0.73 | 12.2 |
| **c-Si(*L >> d*)/a-Si** | 1.6 x 10$^{15}$ | 0.693 | 0.663 | 31.9 | 0.70 | 14.8 |
| **Ref: c-Si(*L >> d*)/metal** | | | 0.550 | 32.2 | 0.73 | 12.9 |

# Supplementary Information

Potential of PEDOT:PSS as a hole selective front contact for silicon heterojunction solar cells

*Sara Jäckle[1,2+], Martin Liebhaber[3+], Clemens Gersmann[3], Mathias Mews[4], Klaus Jäger[5], Silke Christiansen[1,2,6], Klaus Lips[3,6]\**

## S1 Comparison of PH1000 and F HC (Clevios, Heraeus)

For the standard preparation of PEDOT:PSS/c-Si solar cells with a metal back contact we used the polymer solution PH1000 (Heraeus Clevios) mixed with 5 vol% dimethyl sulfoxide (DMSO) and 0.1 vol% of wetting agent (FS31, Capstone), following our previous work.[1,2] The high efficiency PEDOT:PSS/c-Si/a-Si devices were fabricated with the premixed polymer formulation F HC (Heraeus Clevios). While the thickness of the polymer layers depends on the spin coating parameter as well as to a small extend on the substrate size, under similar preparation conditions F HC leads to thinner films than PH1000 (determined by AFM measurements). The specific conductivity of both solutions, measured by the Van de Pauw method on polymer films spin coated on a glass substrate and collected in Table S1, is almost the same. To achieve a comparable polymer film thickness on silicon substrates for solar cells, PH1000 is spin coated at 4000 rpm and F HC at 2000 rpm. Figure S1 shows the photovoltaic response of solar cells fabricated on bulk-limited silicon wafers with metal back contacts featuring the two PEDOT:PSS solutions. The corresponding solar cell parameters are collected in Table S1. Besides a slight decrease in $J_{SC}$, which might be due to a less favorable antireflective behavior of the slightly thinner F HC layer (see also Fig. 4a in the main manuscript), the performance of the solar cell does not differ from the one prepared with PH1000. This shows that these solutions can be used interchangeable when controlling the film thickness.

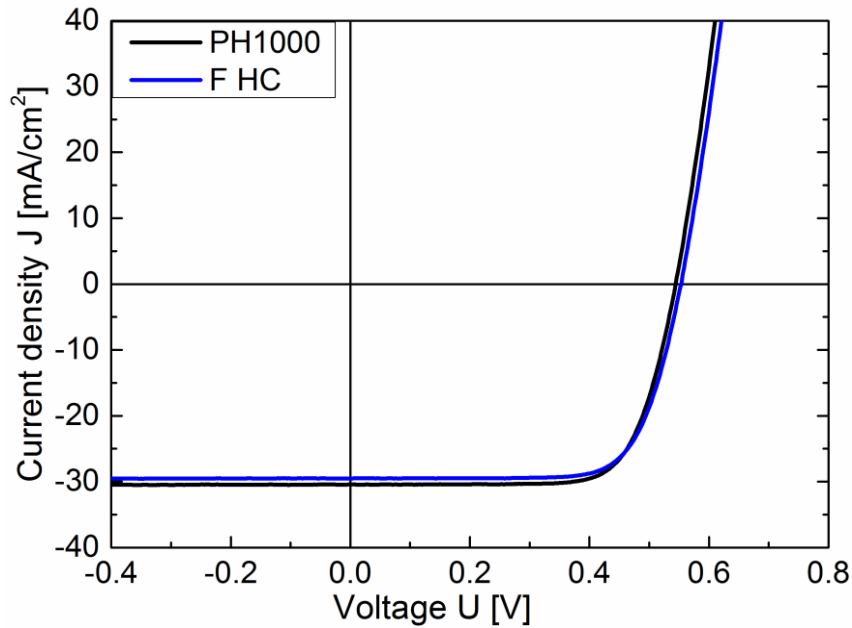

**Figure S1. Photovoltaic response of silicon solar cells with a PEDOT:PSS front junction and a metal back contact** for different PEDOT:PSS solutions. Extracted solar cell parameters are collected in Table S1.

|        | Films (on glass) | | Solar cell parameters | | | | |
|---|---|---|---|---|---|---|---|
|        | d (2k) [nm] | σ [S/cm] | d [nm] | $V_{oc}$ [V] | $J_{sc}$ [mA/cm²] | FF | PCE [%] |
| **PH1000** | ~125 | ~630 | ~85 (4k) | 0.544 | 30.5 | 0.73 | 12.2 |
| **F HC**   | ~70  | ~650 | ~82 (2k) | 0.554 | 29.5 | 0.73 | 11.9 |

**Table S1. Summary of polymer film and solar cell parameters** for different PEDOT:PSS formulations (all abbreviations are defined in the text). Polymer films on glass are spin coated at 2000 rpm, while for solar cell fabrication PH1000 is spin coated at 4000 rpm and FHC at 2000 rpm.

## S2 Optical constants of PEDOT:PSS

The optical constants of the used PEDOT:PSS formulations PH1000 and F HC were derived by spectroscopic ellipsometry (UVISEL, Horiba Jobin Yvon). For the measurements the polymer films were deposited on silicon substrates. Measurements were performed within a photon energy range from 0.6 eV - 4.8 eV and steps of 0.02 eV under an angle of incidence of 70°. The data was analyzed assuming uniaxial anisotropy of PEDOT:PSS, following Pettersson et al.[3] For each component a model consisting of a Lorentz oscillator and a Drude term, accounting for free carrier absorption in the highly doped polymer film, was used.[4] The extracted optical constants in the visible spectrum for the ordinary and extraordinary component are shown in Figure S2.

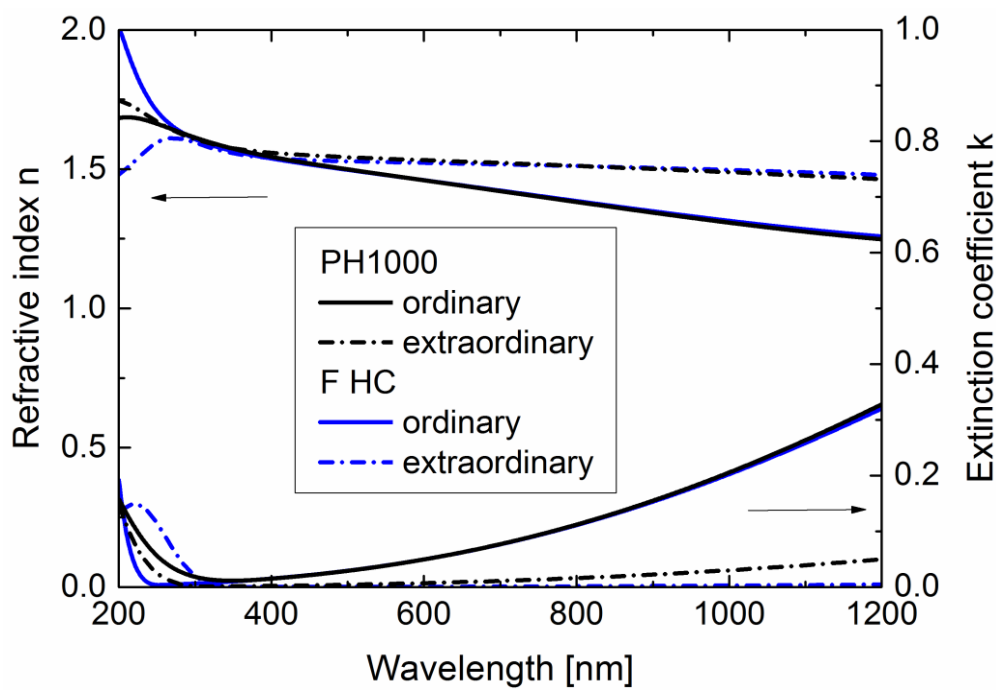

**Figure S2. Optical constants of uniaxial anisotropic PEDOT:PSS measured by ellipsometry** for different PEDOT:PSS formulations.